\documentstyle[11pt,aas2pp4]{article}
\newcounter{currentfig}

\tighten

\slugcomment{To appear in Ap.J.}

\lefthead{Thakar and Ryden}
\righthead{SPH Simulations of Counterrotating Disk Formation}

\def\bea{\begin{eqnarray}}
\def\eea{\end{eqnarray}}
\def\beq{\begin{equation}}
\def\eeq{\end{equation}}

\def\ss{\subsection}
\def\sss{\subsubsection}

\begin{document}

\centerline{(To appear in ApJ)}\vspace*{0.15in}
\title{SPH Simulations of Counterrotating Disk Formation in Spiral Galaxies}
\vspace*{-0.15in}
\author{Aniruddha R. Thakar}
\affil{Center for Astrophysical Sciences, The Johns Hopkins University,
Baltimore, MD 21218-2695; thakar@pha.jhu.edu}
\vspace*{-0.15in}
\author{Barbara S. Ryden\altaffilmark{1}}
\affil{Department of Astronomy, The Ohio State University, Columbus, OH
43210-1106; ryden@astronomy.ohio-state.edu}
\altaffiltext{1}{National Science Foundation Young Investigator.} 

\vspace*{-0.5in}
\begin{abstract}
We present the results of Smoothed Particle Hydrodynamics (SPH) simulations of
the formation of a massive counterrotating disk in a spiral galaxy.  The
current study revisits and extends (with SPH) previous work carried out with
sticky particle gas dynamics, in which adiabatic gas infall and a retrograde
gas-rich dwarf merger were tested as the two most likely processes for
producing such a counterrotating disk.  We report on experiments with a cold
primary similar to our Galaxy, as well as a hot, compact primary modeled after
NGC~4138.  We have also conducted numerical experiments with varying amounts
of prograde gas in the primary disk, and an alternative infall model (a
spherical shell with retrograde angular momentum).  The structure of the
resulting counterrotating disks is dramatically different with SPH.  The disks
we produce are considerably thinner than the primary disks and those produced
with sticky particles.  The time-scales for counterrotating disk formation are
shorter with SPH because the gas loses kinetic energy and angular momentum
more rapidly.  Spiral structure is evident in most of the disks, but an
exponential radial profile is not a natural byproduct of these processes.  The
infalling gas shells that we tested produce counterrotating bulges and rings
rather than disks.  The presence of a considerable amount of preexisting
prograde gas in the primary causes, at least in the absence of star formation,
a rapid inflow of gas to the center and a subsequent hole in the
counterrotating disk.  For a normal counterrotating disk to form, there must
either be little or no preexisting prograde gas in the primary, or its
dissipative influence must be offset by significant star formation activity.
The latter scenario, along with the associated feedback to the ISM, may be
necessary to produce a counterrotating disk similar in scale length and scale
height to the primary disk.  In general, our SPH experiments yield stronger
evidence to suggest that the accretion of massive counterrotating disks drives
the evolution of the host galaxies towards earlier (S0/Sa) Hubble types.

\end{abstract}
\keywords{galaxies: spiral --- galaxies: structure --- galaxies: evolution ---
galaxies: interactions --- galaxies: kinematics and dynamics --- hydrodynamics}

\section{Introduction}
There are only a handful of known cases of {\em massive} counterrotating disks
in spiral galaxies to date, and yet counterrotation in spirals cannot be
deemed a rare phenomenon by any standards.  There are hints that it may be
quite common, in fact, in early-type spirals, particularly S0s (\cite{kfm96}).
The origin of any counterrotating mass (gas or stars) within a spiral disk is
an important unsolved problem with profound implications for the formation
and evolution of all spiral galaxies, but the existence of a significant
retrograde mass component (comprising anywhere from $\sim$10-50\% of the
total mass of the disk system) is a particularly intriguing question that
threatens to radically alter our view of the evolution of spiral galaxies.
The rogues' gallery of spirals with such massive counterrotating disks
currently boasts the following members: NGC~4550 (\cite{rgk92}), NGC~7217
(\cite{mk94}), NGC~4826 (\cite{bwk92}), NGC~3626 (\cite{cbg95}), NGC~3593
(\cite{bcc96}), and NGC~4138 (\cite{jbh96}).

Apart from the challenge they present to the traditional view of spiral
galaxies that has evolved over the last few decades, counterrotating disks
raise several questions about other astrophysical processes, such as the role
and fate of gas in galaxy interactions, star formation in galaxies that
contain counterrotating gas, the accretion rates and star formation histories
of spiral galaxies in general, and the impact of counterrotating populations
on the overall stability of the disk system.  Even though a recent survey of
S0s (\cite{kfm96}) found counterrotating gas in almost a quarter of the
sample, none of these galaxies have counterrotating stars.  Why is the
counterrotating gas not forming stars?  If star formation is inhibited in
counterrotating disks, how does one explain NGC~4550 and others with stellar
counterrotating disks?

It is very unlikely that counterrotating systems can be produced indigenously
or as a byproduct of the galaxy formation process.  The theory of formation of
a spiral galaxy from a spinning protogalactic cloud does not admit the
possibility of bidirectional spin being imparted to the disk system.
Subsequent accretion or merger events are a much more plausible explanation,
and even these are severely constrained by the observed coldness of the
counterrotating galaxies.  Dissipationless mergers, especially between
progenitors with comparable masses, can be ruled out in the general case,
although \cite{pf98} has recently been able to produce a remnant resembling
NGC~4550 with a collisionless merger of two spirals with special initial
conditions.  This leaves minor, gas-rich mergers, and gas accretion or infall,
as the most promising candidates.

The most puzzling aspect of massive counterrotating disks in spiral galaxies
is that the host galaxies appear quite normal in every other respect, and
there is no evidence of excessive thickening of the primordial disks due to
the accretion of the counterrotating disk.  This suggests that the accretion
process must not be a rapid or violent one, but there may be deeper
implications here for the interaction histories of all spiral galaxies.  A
recent study of the tidal thickening of galaxy disks (\cite{rc96}) indicates
that the ratio of scale length to scale height, $h/z_{\circ}$, is 1.5-2 times
lower for interacting disks.  But if any galaxy can be assumed to have
undergone an interaction in the past, the thinness of the non-interacting
sample proves that this ratio returns to its higher value after a certain
amount of time (order of 1 Gyr).  If it can be proved that a spiral can double
its disk mass without destroying the primordial disk in the process, then the
present-day appearance of spiral disks can no longer preclude such
interactions in their past.

The claim that the thinness of spiral disks places a stringent limit on past
accretion (\cite{to92}) is further challenged by recent simulations showing
that the halo absorbs a good portion of the orbital energy and angular
momentum of the satellite in spiral-dwarf mergers (\cite{wmh96}; \cite{hc97}),
and analytical results indicating that an isothermal halo may even shield the
disk from an external tidal field due to a satellite (\cite{mt97}).  A study of
the lopsidedness of the disks of field spirals suggests that the accretion
rate of spiral galaxies may be as high as one small ($\sim$10\% mass)
companion every 4 Gyr (\cite{zr97}).  This is still a very uncertain estimate,
and a more accurate estimate of the accretion rate, and a knowledge of how
much accretion a disk galaxy can withstand, are important questions from a
cosmological point of view.  Observations of counterrotating disks combined
with a good understanding of how they form provide a striking new way to
independently constrain these estimates, because the accreted matter can be
easily distinguished in a counterrotating galaxy.

With the aim of understanding the origin of massive counterrotating disks in
spirals and S0s, we have developed a numerical code and run hydrodynamical
simulations to investigate the processes that are most likely to produce such
bizarre systems.  We have combined an N-body gravity solver with a gas
dynamics particle code for this purpose.  To study the basic parameters of the
processes involved, we first adopted a quick-and-dirty ``sticky particle'' gas
dynamics approach that allowed us to test various scenarios with relatively
small investments of CPU time.  In this way we were able to test massive
counterrotating disk formation in a fiducial cold primary (\cite{tr96},
hereafter TR) as well as model the formation of the recently discovered
counterrotating disk in the early-type spiral NGC~4138 (\cite{tr97}, hereafter
TRJB).  For each type of primary modeled, we have tested two theories of
origin: adiabatic (secular) gas infall and a gas-rich dwarf merger.  In order
to produce a disk with opposite spin, both of these processes require a
retrograde orbit of accretion for the infalling gas or satellite galaxy with
respect to the primary's spin.  Gas infall works well for both late and
early-type primaries, but a dwarf merger, especially with a substantial amount
of dissipationless matter in the dwarf, is not viable for a cold primary
because it plays havoc with the primary's disk.

An accurate representation of the gas in astrophysical systems typically
requires that the hydrodynamical conservation equations be solved and the
effects of physical processes such as shocks and viscosity be included.  This
requires the gas to be modeled as a fluid, but a good compromise can be
achieved with a particulate representation if each gas particle is smeared
over a finite volume and its physical properties averaged or smoothed over
that volume.  This Smoothed Particle Hydrodynamics (SPH) approach
(\cite{lu77}; \cite{gm77}; \cite{mo92}) meshes well with an N-body particle
code, but is much more costly in terms of CPU resources than the sticky
particle approach.  We therefore reserved it for a more detailed and
restricted look at the structure of the counterrotating disk formed by
reexamining (with SPH) a carefully chosen subset of the simulations we
presented in TR and TRJB.  We present the results of these simulations here.
Although most of the simulations are reruns of those presented before with
sticky particles, a couple of new runs are also discussed.  Our SPH code
currently does not include star formation.  We intend to incorporate this in
our simulations in the future, but as we discuss in \S{4} below, we do not
expect it to have a profound impact on the structure of the counterrotating
disks formed.

Our computational method is described in the following section.  Results of
simulations are presented in \S{3}, followed by discussion and conclusions in
\S{4} and \S{5} respectively.

\section{Computational Method}
Our SPH code is an implementation in the C language of the TREESPH code
(\cite{hk89}, hereafter HK).  We started with a basic C version of the tree
code (\cite{bh86}) that performs the force calculation, and optimized it for
execution on the Cray Y-MP at the Ohio Supercomputer Center (see TR for
details).  These changes included rewriting the tree construction and
traversal routines in such a way that they made extensive use of the vector
capabilities of the Y-MP (\cite{he90}; \cite{ma90}).  The outcome was to speed
up the execution by a factor of 5 or more.  We then added the SPH code to it
according to the prescription given by HK, using the same vectorization
techniques as before to optimize the tree traversals required for nearest
neighbor searches.  Vectorization of the SPH calculations was tricky, and we
benefited considerably from a perusal of the vectorized Fortran TREESPH code
(a copy was kindly provided by Lars Hernquist).  To test the final result, we
ran the one-dimensional and 3-d tests described in HK to ensure that it gave
identical results.

For the simulations described here, we have assumed an isothermal gas with a
cutoff temperature of $10^4$ K.  This has been shown to be a fairly good
assumption for the ISM in disk galaxies, where most of the gas hovers close to
the cutoff temperature due to the short time-scale of the radiative cooling
process (\cite{bh91}; \cite{mh94a}, 1994c).  The motivation for assuming an
isothermal gas is that it yields a tremendous saving in the CPU resources
required for each simulation, since solving the energy equation is a non-
trivial exercise that requires a semi-implicit scheme (see HK) and must be
performed at the smallest time-scale involved in the simulation.  

Even with the isothermal assumption, and with the maximum possible
vectorization, our SPH code typically requires 5-10 times the number of CPU
resources required for the sticky particle code.  This is mostly because the
gas particles have their own individual time-steps (each particle's time-step
is a power-of-two subdivision of the system time-step, so that all gas
particles are synchronized at the end of a system time-step as discussed in
HK), and this causes the performance to degrade when there are even a few
particles in the lowest time-step bin.  To keep this problem under control, we
do impose a minimum smoothing length (usually about a quarter of the softening
length), but even so the isothermal assumption is more of a necessity than an
option for us.  As we discuss below, this assumption does present some
problems when we attempt to model a gas-rich dwarf galaxy.

We have adopted the spherically symmetric spline kernel for SPH averaging
(\cite{ml85}, eq. 2.8 in HK).  However, we do not use it for gravity
softening.  Our gravity solver uses the standard softening scheme in which the
gravitational potential is of the form $\Phi\propto(r^2+\epsilon^2)^{-1/2}$
(corresponding to a Plummer density profile), with softening length
$\epsilon=0.5$ kpc for all simulations.  Although our code provides the
capability to use different softening lengths for the collisionless and gas
particles, we have used the same length for both species in the simulations
reported here.  The softening length effectively limits our resolution, but it
is also necessary to suppress the two-body effects in our moderate-$N$
simulations.

Our time-integration scheme is different from the one used by HK.  Their
scheme requires manual synchronization of the particle positions and
velocities by applying midpoint corrections (eqs. 2.42 - 2.45 of HK).  We
decided instead to adopt a slightly different scheme that does not require
these corrections (\cite{kwh96}).  Although this scheme is not time-symmetric
as a result of the particle positions and velocities being advanced in a
different way than in HK, it is simpler and faster without introducing any
additional error.

For all simulations reported here, we use the combined form of the bulk and
Von Neuman-Richtmyer artificial viscosity given by eqs. 2.24-2.25 in HK, and
applied the restriction necessary to avoid viscous cooling of receding
particles as specified by HK.  This viscosity introduces less shear than other
artificial viscosities, although it does not provide as accurate a description
of the flow near shocks.  We felt that it was more important to avoid
excessive shear in disk galaxy simulations, and therefore adopted the combined
form of viscosity.

We use 32k particles each to represent the primary's halo and disk, and up to
20k particles for the gas depending on its mass.  Where a dwarf galaxy is
involved, the stars and dark matter in the dwarf are represented by enough
particles so that the per particle mass is roughly the same as in the case of
the primary.  A nagging problem with disk simulations is the disk heating due
to the graininess of the potential that is inherent in a particle code
(\cite{he89}).  The problem is more severe in our simulations because of the
longer time-scales involved in retrograde interactions.  Even with a softening
length as high as 0.5 kpc, we find that there is considerable disk heating
over the duration of our simulations with the number of particles we have
chosen, and the heating of the disk due to the interaction itself is always of
the order of the numerical heating.  Unfortunately, the two are not separable,
and this limits our ability to estimate the disk thickening accurately.
Although we would like to use more than 32k particles each to represent the
disk and halo, our CPU resource limitations currently prevent us from running
considerably larger N simulations.

\begin{table*}[!htb]
\caption{Parameters for Primary Galaxy.}
\begin{tabular}{lcccccccc}
\\ \tableline\tableline
& \multicolumn{4}{c}{Disk} & & \multicolumn{3}{c}{Halo} \\
\cline{2-5} \cline{7-9} 
& Mass & Radius & Scale Len. & Scale Ht. & & Mass & Radius & Core Rad. \\ 
& $M_d$ [$M_{\sun}$] & $R_D$ [kpc] & $R_d$ [kpc] & $h_d$ [kpc] & & $M_h$
[$M_{\sun}$] & $R_h$ [kpc] & $R_c$ [kpc] \\ \tableline
Cold Primary [TR] & $5.5\times10^{10}$ & 21.0 & 3.50 & 0.3 & & $2.2\times10^{11}$ & 50.0 & 1.0 \\
NGC~4138 [TRJB] & $2.0\times10^{10}$ & 6.0 & 1.25 & 0.2 & & $1.5\times10^{11}$ & 20.0 & 5.0 \\
\tableline
\end{tabular}
\label{ppars}
\end{table*}

The primary galaxy physical parameters are the same as we used before for the
cold primary (TR) as well as for NGC~4138, the hot primary (TRJB).  For
convenience, they are repeated in Table~\ref{ppars}.  Our fiducial cold
primary parameters are typical for an Sbc spiral like the Milky Way, and the
disk is bar-unstable as is consistent with its coldness (\cite{bt87}).  Unless
explicitly stated, the primary disk in most of the simulations reported here
contains no primordial gas.  The halo is a truncated isothermal sphere
(truncated at $r=R_h$) with a logarithmic profile and core radius $R_c$ as per
eq. 2-54a in Binney \& Tremaine (1987).  Although most of the SPH simulations
are reruns of previous sticky particle runs, there are some changes and some
new simulations.  The input parameters for the gas infall and merger
simulations are listed in Tables~\ref{ipars} and \ref{mpars} respectively.
Correspondences to the original simulations in TR and TRJB are pointed out,
where applicable, when we discuss the results in the following section.  To
avoid confusion with the sticky particle runs, we have named the SPH gas
infall runs G1-G8, and the SPH dwarf merger runs D1 and D2.

The models for gas infall and mergers are the same as in TRJB.  The infalling
gas is configured as a long rectangular column of uniform density and square
cross-section.  The gas columns we use in the SPH simulations may be shorter
than the corresponding sticky particle simulations, primarily because the SPH
runs require much longer to run and hence a shorter column with a
correspondingly shorter infall time is preferable.  The primary disk plane is
selected to be the $xy$-plane, and the long axis of the infalling gas slab is
parallel to the $y$-axis.  The primary disk rotation is counter-clockwise, and
the center of the primary disk is at the origin of the coordinate system.  The
gas is then placed in the first quadrant in the same plane as the primary
disk, and is given an initial kick mostly in the negative $y$ direction (an
initial speed in the positive $x$ direction is also given in order to increase
the initial angular momentum of the gas in some cases, see Table~\ref{ipars}).
This ensures that the gas is accreted on a retrograde orbit with respect to
the primary disk spin.

\begin{table}[!htb]
\caption{Parameters for Gas Infall Simulations.}
\begin{tabular}{lccccccc}
\\ \tableline\tableline
 & $M_g\tablenotemark{a}$ & $V_{0y}\tablenotemark{b}$ &
$V_{0x}\tablenotemark{c}$ & $Y_0\tablenotemark{d}$ & $X_0\tablenotemark{e}$ &
$L_0\tablenotemark{f}$ & $T_0\tablenotemark{g}$ \\ \tableline
G1 & $4.40$ & 0.3 & 0.10 & 50 & 50 & 300 & 20 \\
G2 & $4.40$ & 0.3 & 0.00 & 50 & 50 & 300 & 10 \\
G3 & $1.32$ & 0.3 & 0.05 & 25 & 25 & 150 & 10 \\
G4 & $1.32$ & 0.3 & 0.05 & 25 &  5 & 100 & 10 \\
G5 & $1.45$\tablenotemark{\dag} & 0.3 & 0.10 & 25 & 10 & 100 & 10 \\
G6 & $1.76$\tablenotemark{\dag} & 0.3 & 0.05 & 25 &  5 & 100 & 10 \\
G7 & $1.06$\tablenotemark{\ddag} & 0.2 & - & - & 30 & - & 10 \\ 
G8 & $1.06$\tablenotemark{\ddag} & 0.1 & - & - & 30 & - & 10 \\ \tableline
\end{tabular}
\label{ipars}
\tablenotetext{\rm a}{total gas mass [$10^{10} M_{\sun}$]}
\tablenotetext{\rm b}{initial velocity of the gas in the negative $y$
direction, specified in units of the centripetal velocity 
$\sqrt{G(M_h+M_d)/X_0}$, where $M_h+M_d$ is the total mass of the primary
(halo+disk)}
\tablenotetext{\rm c}{initial velocity of the gas in the positive $x$
direction, specified as a fraction of $V_{0y}$}
\tablenotetext{\rm d}{initial $y$-distance of the gas column from the center
of the primary disk [kpc]}
\tablenotetext{\rm e}{initial $x$-distance of the gas column from the center
of the primary disk [kpc]}
\tablenotetext{\rm f}{length of the infalling gas column [kpc]}
\tablenotetext{\rm g}{thickness of the gas column (side of square cross
section) [kpc]}
\tablenotetext{\dag}{G5 includes a prograde ring of gas in the primary with
$\sim10$\% of the mass of the infalling gas (hence $M_g=1.32+0.13$), and G6
includes a prograde disk of gas in the primary with $\sim20$\% of the mass of
the stellar primary disk (hence $M_g=1.32+0.44$)}
\tablenotetext{\ddag}{G7 and G8 are infalling gas shells having retrograde
angular momentum with rotational speed $V_{0y}$, radius $X_0$ and thickness
$T_0$.}
\end{table}

\begin{table}[!htb]
\caption{Dwarf Galaxy Parameters for Mergers.}
\begin{tabular}{lccccccc}
\\ \tableline\tableline
& $M_g$\tablenotemark{a} & $M_*$\tablenotemark{b} &
$M_D$\tablenotemark{c} & $F_g$\tablenotemark{d} & $R_*$\tablenotemark{e}
& $R_g$\tablenotemark{f} & $R_h$\tablenotemark{g} \\ 
\cline{2-4} \cline{6-8}
& \multicolumn{3}{c}{[$10^{10} M_{\sun}$]} & & \multicolumn{3}{c}{[kpc]} \\
\tableline  
D1/D2 & 1.06 & 0.26 & 2.64 & 0.27 & 4.0 & 4.0 & 8.0 \\ \tableline
\end{tabular}
\label{mpars}
\tablenotetext{\rm a}{mass of gas}
\tablenotetext{\rm b}{mass of stars}
\tablenotetext{\rm c}{mass of dark matter}
\tablenotetext{\rm d}{ratio of gas mass to total mass}
\tablenotetext{\rm e}{radius of stellar sphere}
\tablenotetext{\rm f}{radius of gas sphere}
\tablenotetext{\rm g}{radius of halo}
\end{table}

For the dwarf mergers, the setup is basically the same except that the dwarf
galaxy replaces the gas column in the first quadrant.  The dwarf galaxy model
is the same as described in TR and TRJB.  We have only attempted two gas-rich
dwarf mergers (D1 and D2), both with NGC~4138 as the primary, since they were
not feasible for the cold primary due to the problems discussed in TR.  Both
mergers have identical input parameters and are a repetition of merger M2 from
TRJB.  SPH is used from the beginning for D1, whereas for D2 the merger is run
with sticky particles until the dwarf galaxy is completely disrupted by the
primary, at which point SPH replaces sticky particles for further
calculations.  This is due to the difficulties we encountered in using
isothermal SPH to model the dwarf galaxy (see \S3.2.2).

\section{Results}

The primary galaxy is evolved in isolation for a few dynamical times before
the gas or dwarf galaxy is introduced.  Hence the simulation results are shown
starting at $t\gtrsim1.5$ Gyr for the cold primary, and $t\gtrsim0.5$ Gyr for
NGC~4138 (which has a smaller dynamical time).  The simulation stops once most
of the gas has settled in the plane of the primary and the evolution of the
gas disk has tapered off to nearly zero.
 
\ss{Fiducial Cold Primary} 

For the cold primary, our first infall simulation is a repetition of the
second continuous infall simulation discussed in TR, but in the SPH run we use
a shorter column of gas.  The infall simulation shown in Fig.~\ref{g1gas} has
all other inputs the same as before (see Table~\ref{ipars} for input parameter
values).  A counterrotating disk with spiral structure has started to form by
$t\sim4.5$ Gyr.  The side view confirms the disk formation and indicates that
the counterrotating disk is uniformly thin throughout.  Since the width of the
last two panels in Fig.~\ref{g1gas} is 40 kpc, the size of the counterrotating
disk appears to be comparable to that of the primary (radius 21 kpc).  One has
to be careful in making this comparison, however, since the radial profile of
the counterrotating disk is not close to that of the primary.  Saying that the
sizes are comparable does not mean that the scale lengths are comparable.  The
lack of a centrally concentrated or exponential radial profile is clear from
both the side and top views for $t\gtrsim5$ Gyr.  There is a hole in the inner
regions of the counterrotating disk, and this, along with the larger size,
could be the result of the gas having too much angular momentum initially.

Although the observed instances of counterrotating disks do not show a
consistent radial profile, it is nonetheless important to determine whether a
counterrotating disk with a similar scale-length and profile as the primary
disk can be formed.  There is at least one example of this, NGC~4550
(\cite{rfi92}).  We have experimented with lower values of the initial angular
momentum of the infalling gas to determine what effect it has on the radial
profile and structure of the counterrotating disk formed.  Fig.~\ref{g2gas}
presents the results of run G2, which has lower initial angular momentum of
the gas as well as a narrower (and hence denser) column of infalling gas than
G1.  Since the panel widths in Fig.~\ref{g2gas} are the same as in
Fig.~\ref{g1gas}, the differences are easy to see.  The size and structure of
the counterrotating disk formed are quite different.  The disk is smaller (by
about 25-30\%) than the disk formed in G1, and the radial profile is more
centrally concentrated.  There is a lot more structure visible as the gas has
piled up in several places due to shocks.  The side view shows that the disk
is as thin as in G1, and it is significantly thinner than the primary disk.

A comparison of the primary disk at the beginning and the end of G2 with the
counterrotating disk is shown in Fig.~\ref{g2disks}.  The primary disk shows a
bar even before the gas is introduced, but the bar does not become more
pronounced after the formation of the counterrotating disk.  It is, in fact,
obscured by the inclination of the disk in the top view of the final timestep.
The final thickness of the primary disk is hard to measure because of its
inclination, but we do measure it by compensating for the inclination and
computing the $z=0$ plane at each location in the disk (see below).  However,
it is obvious from Fig.~\ref{g2disks} that the primary disk thickness has more
than doubled, and it now resembles a lenticular (S0) disk rather than an Sb
disk.  This is further corroborated by the rotation curves and velocity
dispersion plots for the final disks shown in \S4, where we also discuss some
of the possible reasons for the pronounced heating experienced by the cold
primary.

The counterrotating disk is of comparable size but is less than half as thick
as the primary disk.  The velocity fields of the primary disk and the
counterrotating disk are shown in Fig.~\ref{g2vf}.  The primary's velocity
field is considerably hotter than the velocity field of the counterrotating
disk, but they are clearly antiparallel.

The radial profiles of the primary disk and the counterrotating disks at the
end of simulations G1 and G2 are compared with each other in
Fig.~\ref{g12rad}.  The dip in the particle density between $\sim1-3$ kpc is
clearly visible for G1, and although there is a slight dip at those radii even
for G2, the G2 profile comes much closer to the exponential radial profile of
the primary disk.

The thickness of the primary disk is plotted for simulations G1 and G2 in
Fig.~\ref{g12zth}.  A comparison of the thickness plots with those in TR shows
some similarities.  The thickness of the post-accretion primary disk is only
slightly higher than the thickness of the isolated primary disk evolved over
the same time period.  The actual thicknesses for G1 and G2, however, are
significantly higher (by about 25\%) than the thickness plotted for continuous
infall in TR.  This is probably due to the fact that we are using a smaller
softening length (0.5 kpc instead of 1.0 kpc used by TR), even though we have
doubled the number of particles used to represent the halo for the simulations
in this paper (we used 16k particles for the halo in TR).  The thickness curve
for the isolated disk is also higher by a similar amount compared to the
corresponding curve in TR.  Some of the increase in the heating of the primary
can also be attributed to the fact that the infalling gas column is shorter
and hence more dense than the one used in TR, and the time-scale for formation
of the counterrotating disk has decreased by $\gtrsim30$\% (see \S{4.1}).
This gives the primary disk less time to adjust.  Furthermore, the smaller
softening length aggravates the gravitational impact of the incoming gas on
the primary.

We are not so concerned with the heating of the primary in the simulations
reported here, because we have established in TR and TRJB that if the rate of
accretion is low enough, the effect on the primary can be kept to an
acceptable level.  As mentioned above, our motivation for hastening the
counterrotating disk formation process by using a shorter column of gas than
before was to save CPU time expended on each simulation.  The fact that this
causes considerably more heating of the primary underlines the sensitivity of
the cold primary to the rate of accretion.

\ss{Hot Primary - NGC~4138} 
\sss{Gas Infall} 

The first infall simulation for NGC~4138, G3, is a repetition of simulation G4
from TRJB, although the total mass of the gas is 25\% higher in the SPH runs.
The remaining parameters are the same as in the sticky particle run.  The
results are shown in Fig.~\ref{g3gas}.  The counterrotating disk has started
to form at $t\sim1.8$ Gyr, and after $t\sim2.6$ Gyr, it does not evolve very
much.  The size of the counterrotating disk is significantly larger than the
primary since it extends over more than two-thirds of the width of the panel
(20 kpc) in the side-view at $t\geq2.6$ Gyr, but it clearly does not have an
exponential radial profile.  There is a small central mass concentration.  The
lack of an exponential radial profile is easier to see in the side view at
$t=3.0$ Gyr, because the disk is slightly tilted to the line of sight, and an
outer ring is visible in addition to the central concentration, with a drop in
the particle density at intermediate radii.  The radial density distribution
is plotted below in Fig.~\ref{g34rad}.  As in the case of the cold primary
(G1), the lack of an exponential profile indicates that a good fraction of the
particles do not lose enough angular momentum to allow them to settle in the
inner few kpcs of the disk.  Spiral structure is evident but it is not very
strong.  The side view at $t=2.6$ Gyr shows the counterrotating disk to be
quite thin, although this is hard to see in the last panel because of the
inclination of the disk.

The velocity fields of the primary and counterrotating disk are compared in
Fig.~\ref{g3vf}.  The larger size of the counterrotating disk is also evident
in the velocity field plot.  The velocity field of the counterrotating disk is
colder than that of the primary and more regular.

A lower angular momentum version of G3 is attempted in G4, shown in
Fig.~\ref{g4gas}.  The initial velocity of the gas column is the same as G3,
but the initial distance of the gas from the center of the primary is smaller
by a factor of 5 (see Table~\ref{ipars}).  The length of the gas column is
also two-thirds of that in G3.  The parallels between G1/G2 and G3/G4 are
quite obvious.  Once again, the lower angular momentum produces a
counterrotating disk with a more centrally concentrated radial profile and a
smaller overall size.  A comparison of the radial profiles of G3 and G4 with
each other and the primary's radial profile is shown in Fig.~\ref{g34rad}.
Even though the radial profile of the counterrotating disk in G4 is not very
close to the exponential profile of the primary, it is much closer to that
than G3.  There is no pronounced dip in the particle density in the inner
radii for G4, as there is for G3.

There is not much difference in the heating (and thickening) of the primary
disk with the SPH simulations of NGC~4138 compared to TRJB.  Although the
heating is a little higher on average in our SPH simulations, this is due to
the fact that we are using shorter gas columns.  This makes the accretion rate
higher.  In the cases where a lot of gas does make it to the nuclear portions
of the primary, there is slightly more heating of the primary.  In the other
cases, the levels are comparable to those obtained in TRJB (after allowing for
the shorter gas columns), and hence the thickness plots are not repeated here.

\sss{Gas-Rich Dwarf Merger}

Our previous studies have established that the gas-rich dwarf merger model of
counterrotating disk formation is only feasible for a hot primary like
NGC~4138.  TR found that even a merger with a 10\% mass satellite produced
unacceptable levels of heating in a cold primary disk, whereas the hot primary
used by TRJB fared much better in this regard and was able to withstand a
gas-rich dwarf merger under restricted conditions.

Apart from the issue of whether a dwarf merger can produce a counterrotating
disk, we encountered a technical difficulty with our gas-rich dwarf galaxy
model.  A stable model of a collisionally supported gas-rich dwarf galaxy is
not easy to obtain with an isothermal equation of state, because the gas in
the dwarf tends to shock and collapse rapidly if the temperature is too low,
and expand if it is too high.  This inability of the isothermal model to yield
a stable configuration is not very surprising, since isothermal pressure
gradients do not grow rapidly enough to combat the gravitational collapse of
the gas (\cite{to80}).  A rotationally supported dwarf would probably not
alleviate this situation very much, because it would still collapse in the
direction not affected by the centrifugal force, thereby yielding a
quasi-equilibrium isothermal disk.  As such, we restrict ourselves to testing
only one model for which we have a reasonable expectation of success, and even
for this we have to adjust the gas temperature to keep the dwarf galaxy in
equilibrium.  

We have selected merger M2 from TRJB, with a dwarf that is $\sim18$\% as
massive as the primary.  Two versions of this merger are presented here.  In
the first case (run D1), we use SPH with a higher isothermal temperature until
the dwarf galaxy is tidally disrupted by the primary, and thereafter reduce
the temperature of the gas to the usual value ($10^4$ K).  In the second
version (run D2), we use the sticky particle results to evolve the dwarf until
it forms a thick disk around the primary, and then switch to SPH with the
usual settings.

The results for D1, with SPH being used all the way, are shown in
Fig.~\ref{d1gas}.  The top view shows that the dwarf makes one entire pass
around the primary before getting completely stripped, and ends up forming a
very barred counterrotating disk.  In the side view the disk appears to be
quite thin, and the velocity field diagrams shown in Fig.~\ref{d1vf} confirm
that a disk is indeed formed.  The stellar particles (not shown) of the dwarf
form a thick flattened cloud around the primary, in the same way as they did
with sticky particles.  The velocity fields also indicate that the primary
disk is heated up more than in the infall case (compare with Fig.~\ref{g3vf}).

When SPH is introduced at a late stage after the dwarf has already been
tidally stripped and has formed a thick disk around the primary, the evolution
of the disk is very different (Fig.~\ref{d2gas}).  D2 dramatically illustrates
the differences between sticky particle gas dynamics and SPH.  Within a short
time ($\lesssim0.4$ Gyr), the thick sticky particle disk has collapsed to a
thin disk.  There is a slow accumulation of mass in the center thereafter, but
beyond the formation and dissipation of a few rings, the disk is fairly
stable.

It appears that the nature of the dissipation is very different with SPH.
With sticky particles, there is a slow but steady loss of kinetic energy,
whereas with SPH, shocking causes rapid loss of energy.  The isothermal model
probably contributes to the speed of the energy loss, since there is no
counteracting increase in thermal pressure to oppose the kinetic energy
dissipation until the cutoff temperature is reached.  Once the SPH gas forms a
kinematically cold disk, evolution ceases as the dissipation drops off
suddenly.  The rapid loss of kinetic energy is demonstrated even more
effectively when prograde gas is present in the primary (see below) and the
counterrotating gas collides head-on with it.

\ss{Other Experiments with NGC~4138}

\sss{Prograde Gas in the Primary}

Sticky particle simulation I6 (TRJB), which included a ring of prograde gas in
the primary, showed that the prograde gas did not have much of an impact on
the kinematics of the counterrotating gas.  Collisions between
counterstreaming gas particles did send some gas to the center, but this was a
modest effect and the prograde gas ring remained mostly intact.

The SPH simulation G5 shown in Figs.~\ref{g5gas} and \ref{g5vf} is a
repetition of I6 from TRJB.  The average radius of the ring is 1.7 kpc and the
width of the ring is 1 kpc (inner radius of ring = 1.2 kpc and outer radius
= 2.2 kpc).  The mass of the ring is $\sim10$\% of the mass of the infalling
gas, and the ring has uniform density.  The particles in the ring are given
initial velocities appropriate for primary disk particles at those radii, and
the ring is checked for stability for a couple of dynamical times prior to
introducing the infalling gas.  

We find that the behavior of the gas in the SPH version of this simulation is
dramatically different from the sticky particle results.  Virtually all of the
prograde gas interacts with the counterrotating gas and falls to the center,
leaving a gap in the counterrotating gas distribution at the original location
of the prograde ring of gas.  The interaction between the prograde and
retrograde gas particles can be seen more clearly in the velocity fields plot.
The angular momentum of the retrograde gas is reversed near the center of the
primary as both the prograde and retrograde gas particles fall to the center.

Going one step further, we included a prograde {\em disk} of gas in the
primary along with the stellar disk.  The mass of the gas disk is $\sim20$\%
of the mass of the primary disk, and the radius of the gas disk is 4 kpc.  As
was the case for the prograde ring, the prograde disk is of uniform density,
and we evolved it with the rest of the primary first for 200 Myr (without any
counterrotating gas) to ensure that it was stable on its own.  Then we
introduced the infalling gas at $t=0.7$ Gyr.  This is simulation G6 shown in
Fig.~\ref{g6gas}.  The behavior of the two gas components is similar to that
seen in G5, only stronger.  The prograde gas sweeps up all the retrograde gas
that it comes in contact with, and both combine to form a nuclear mass
concentration within a Gyr or so.  As in the case of the prograde ring, the
velocity field plots (Fig. ~\ref{g6vf}) show that the angular momentum of the
infalling gas is reversed in the process.  The rest of the infalling gas that
does not come into contact with any of the prograde gas forms a
counterrotating outer ring.  The gap between the prograde nuclear disk/bulge
and the outer counterrotating disk is larger in this case since the prograde
gas occupied a larger area in G6 than in G5.

\sss{Infalling Spherical Gas Shell} 

Gas infall is not well-constrained by observations, and as an alternative gas
infall model, we tried a spherical shell of gas that is given retrograde
angular momentum with respect to the primary disk.  The two runs with such an
infalling gas shell, G7 and G8, have the input parameters shown in
Table~\ref{ipars}.  G8 has half the angular momentum of G7.  The results for
G7 and G8 are shown in Figs.~\ref{g7gas} and \ref{g8gas}.  The shell collapses
to a flat disk within a very short time ($\lesssim 0.5$ Gyr).  The speed of
the collapse is somewhat higher for G8.  The final size of the disk in G8 is
about half the size of the disk in G7 (note that the last two panels of
Fig.~\ref{g8gas} are half as wide as the corresponding panels in
Fig.~\ref{g7gas}).  The mass distribution of the two counterrotating disks is
also different.  There is an outer ring present in both disks, but only G8
shows a significant central mass concentration.  The tilting of the disk at
the end of G8 is in phase with the tilting of the primary disk.  The formation
of the outer ring in both cases, together with the dependence of the size of
the ring on the initial angular momentum, suggests that the infalling shell
mechanism is not particularly suitable for counterrotating disk formation.
There is no a priori reason to believe that an infalling gas shell is
physically more realistic, and it certainly does not offer any advantages over
the infalling gas column that we have chosen for most of our infall
experiments.

The formation of rings in the collapse of the spherical gas shells is not a
big surprise.  Ring formation in the collapse of rotating spherical systems
has been documented by several workers, with transient rings forming in
dissipationless collapses (\cite{ms79}), and more stable rings forming in the
collapse of rotating isothermal gas clouds (\cite{to80}; \cite{bh82}).  The
equilibrium of self-gravitating isothermal gas rings has been investigated
from a theoretical perspective as well (\cite{os64}).  Unlike the rings seen
in dissipationless collapses (\cite{ms79}), the rings we see are not transient
structures that are formed en route to a more permanent morphology.
Furthermore, they are true rings in that they are not due to orbit crowding,
and the particles in them are not constantly moving in and out of the ring
configuration.  Our rings are consistent with those seen and studied in
isothermal gas cloud collapse scenarios, and are self-gravitating: the mass
per unit length is much greater than the equilibrium mass per unit length for
isothermal rings (\cite{os64}, eq. 156).  For G7, this quantity is $\sim4$
times, and for G8, it is $\sim7$ times the equilibrium value at the last
time-step, which corresponds to $\sim11 t_{\rm ff}$ for G7 and $\sim8 t_{\rm
ff}$ for G8 (where $t_{\rm ff}$ is the free-fall time of the initial spherical
shell).  However, collapse is most likely inhibited because our numerical
resolution (as set by the softening length, which is 0.5 kpc) is of the order
of the ring width.  Our values of $\alpha$ and $\beta$, the ratios of the
thermal and rotational energies to the gravitational energy respectively
(\cite{bh82}), are $0.02$ and $0.12$ for G7, and $0.02$ and $0.07$ for G8.
The value of $\alpha$ is well below the maximum value that will produce a
collapse.  For G7, the $\beta$ value falls within, whereas for G8, the $\beta$
falls below the $0.1-0.5$ range seen by Boss \& Haber (1982) for isothermal
rings.  However, a lower $\beta$ limit for isothermal rings has not been
established as yet, and it should be remembered that our initial configuration
is a spherical shell rather than a cloud.

\section{Discussion}

\ss{SPH vs. Sticky Particles}

Our primary aim with sticky particle simulations (TR and TRJB) was to
determine the range of input parameters that worked, and to investigate the
effect of the infall/merger on the primary disk.  We succeeded in achieving
these objectives, and we did not put much faith in the sticky particle results
where the structure of the counterrotating disk was concerned.  The SPH
simulations presented here confirm that we were justified in our mistrust of
the sticky particle results.  Shocks and viscosity paint quite a different
picture of counterrotating disk structure.
 
The most visible differences in the disks formed with SPH are their thinness
and radial structure.  Sizes are comparable to the sticky particle disks,
although the SPH disks do on average tend to be a little smaller.  However,
they are all significantly thinner than their sticky particle counterparts.
We found the sticky particle disks to be on average thicker than the primary
disks.  With SPH the reverse is true.  The velocity field diagrams also show
that all the counterrotating disks are considerably colder than the primary
disks.  This was true even with the sticky particle disks.

There is more structure visible in the SPH disks.  A comparison of the final
panels in Figs.~\ref{g1gas}, \ref{g2gas}, \ref{g3gas} and \ref{g4gas} with
Figs.~1, 4, 6 (except 7) of TR and Figs.~2-4 and 8 of TRJB illustrates the
differences in the nature of the SPH disks.  None of the disks produced with
sticky particles had radial profiles approaching an exponential mass
distribution.  We were able to come close with SPH (runs G2 and G4).  The
counterrotating disk produced in G4 shows clear evidence of lopsidedness
(Fig.~\ref{g4disks}), and warps are seen in nearly all of the thin disks
formed with SPH.

The time-scale of the counterrotating disk formation is significantly smaller
(by $\gtrsim30$\%) with SPH.  A comparison of simulations G1-G4 with their
sticky particle counterparts proves this.  The sticky particle run
corresponding to G1 produced a counterrotating disk in $\gtrsim6$ Gyr (TR),
whereas G1 and G2 take $\gtrsim4$ Gyr.  For NGC~4138, sticky particle
simulations were able to produce a counterrotating disk in $\gtrsim3$ Gyr
(TRJB), whereas G3 and G4 achieve the same result in $\gtrsim2$ Gyr.  Most of
the decrease in the time required is due to the shorter columns of gas used in
the SPH simulations, but a comparison of the evolution of the sticky particle
(we ran a sticky particle simulation with the same inputs for comparison) and
SPH gas also shows that the latter loses its kinetic energy and angular
momentum more rapidly.  We estimate that the difference in the time-scales due
to the gas dynamics alone is of the order of 5-10\%.  The infalling gas is
captured and assimilated faster into the primary disk because of the more
efficient viscous dissipation in SPH.

Strangely enough, the infalling gas does not collapse and fragment like it did
with sticky particles.  The basic character of dissipation appears to be
different: in sticky particle dissipation, head-on collisions were not quite
so dissipative, and there was a comparable amount of dissipation regardless of
how much collisional support the gas had; in SPH, whenever collisions between
gas particles cause shocks, the dissipation is drastic and the gas quickly
loses most of its collisional kinetic energy.  As a result, as long as the SPH
gas motion does not involve any head-on collisions between opposing gas
streams, the gas dissipates very little energy.  The tendency of SPH to cause
rapid loss of kinetic energy and angular momentum in counterstreaming gas has
been noted by other users of SPH as well (\cite{mi97}), and it is possible
that SPH is too dissipative in nature (the artificial viscosity too high)
under these circumstances.

\ss{Source of Infalling Material}

While it is fine to theorize about the formation of counterrotating disks by
gas infall or even gas-rich dwarf infall, the evidence for such material in the
intergalactic medium is still quite thin, which is the main reason why there
are no realistic models of infall available.  The possibility of undetected
molecular gas in the IGM around spirals provides one glimmer of hope.  There
has been more evidence recently for a significant fraction of the neutral ISM
in our galaxy being $H_2$-bearing molecular clouds (\cite{dhb98}).  The
general case for ongoing gas infall in spiral galaxies is also getting
stronger, with recent evidence from IRAS data (albeit still somewhat
controversial), that massive star formation rates are independent of spiral
type and that the median gas recycling time is $\sim1\times10^9$ yr
(\cite{dh97}), pointing to a sustainable outside source of gas.

``Galaxy harassment'' has been recently suggested as a means of spiral
evolution in clusters (\cite{mk96}).  Galaxy harassment refers to multiple,
high-speed encounters that transform small spirals into dE/dSph, but more
importantly, leave giant tidal debris arcs and tails that could provide fuel
for quasars as well as the raw material for producing counterrotating disks.
The recent discovery of a giant debris arc in the Coma cluster (\cite{tm97})
supports this theory.  Harassment of the debris tails themselves would create
tidal shocks leading to formation of dwarf galaxies in the tidal tails, as
shown also in numerical simulations (\cite{bh92}).  This would mean that the
incidence of counterrotating disks in clusters should be higher than in the
field, something that can be tested for in the future as more systematic
surveys of counterrotation in clusters (especially the more distant ones) are
undertaken.

The main difficulty posed by dwarf mergers is the heating experienced by the
primary disk due to the large amounts of dark matter currently believed to be
associated with most dwarf galaxies.  There is still some doubt whether the
evidence for the dark matter in dwarf galaxies is trustworthy or not, and if
vindicated, dwarf spheroidal satellites without dark matter (\cite{kr97}) may
provide a better chance of obtaining counterrotating disks from dwarf mergers.
The instances of counterrotating bulges in spiral galaxies (e.g. NGC~7331,
\cite{pgp96}) and counterrotating cores in ellipticals certainly point to
merger events (\cite{bq90}).  We are able to create a counterrotating bulge
with our gas-rich dwarf model when the dwarf is sufficiently dense and massive
(simulations S1 and S2 in TR), but the dissipationless (dark and stellar)
matter in the dwarf is a severe liability to the survival of the primary disk.

\ss{Gravitational Influence of Various Components}

To compare the gravitational influence of the three components (halo, disk and
gas) in the disk plane, we have plotted the gravitational force experienced by
a test particle of unit mass in the mid-plane of the disk ($z=0$) as a function
of radius in Fig.~\ref{fgrav}.  This is shown at the beginning and end of the
simulation, for simulations G2 (cold primary) and G4 (hot primary).  The
gravitational forces experienced by the test particle due to the halo
particles, the primary disk particles, and the gas particles, are shown
separately.  The forces are further separated into their azimuthally averaged
radial (in plane of disk) and vertical components.  Values for the outer
halves of the disks are less reliable due to low particle numbers, and for G2
the calculation of the force components is further complicated by the
substantial inclination of the primary and gas disks at the end of the
simulation.  We have attempted to correct for this inclination as best we can
by first applying a coordinate transformation to reset the disk plane to the
$xy$-plane before doing the force calculations.

For the cold primary (G2), initially ($t=1.5$ Gyr) the primary disk exerts the
dominant radial force in the inner half of the disk system.  The disk and halo
radial forces are almost equal in the outer half.  The vertical forces due to
all components are initially an order of magnitude lower than the radial
forces.  As expected, the influence of the infalling gas is negligible
initially.  This is because the gas is several disk radii away at the
beginning of each simulation.  At the end of the simulation ($t=6.0$ Gyr), the
halo-disk force distribution with radius does not change significantly, but
the magnitude of the forces is lower on average by $\sim7$\%.  The radial force
due to the gas is now comparable to the halo and disk radial forces.  In the
first couple of kpc, the gas radial force dominates.  The vertical influence
of the halo has increased and that of the primary disk has diminished by a
considerable amount ($\sim50-60$\%), especially in the outer half of the disk
plane.  The gas now exerts a vertical force that is dominant in the inner half
of the disk, and is intermediate between the halo and disk vertical forces in
the outer half.  On average, though, the vertical forces are still an order of
magnitude lower than the radial forces.

For the hot primary (G4), the initial ($t=0.5$ Gyr) breakdowns are similar
although the halo is even less dominant in the inner half of the disk plane
due to its high core radius (5 kpc).  The vertical forces are as before an
order of magnitude lower than the radial forces, and the gas influence is
negligible initially.  At the end of the simulation ($t=2.5$ Gyr), the
influence of the gas is close to that of the halo in the inner regions, but is
significantly less than the disk and halo forces in the outer disk.  The
vertical influence of the halo remains much the same, whereas the vertical
force due to the primary disk increases in the inner half.  The gas vertical
force rivals that due to the halo throughout, and is not the dominant vertical
influence as was the case for the cold primary.  The gas mass fraction
(compared to the primary disk mass) is much less in the hot primary, and so it
is not surprising that the gas has a smaller influence both radially and
vertically.

\ss{The Nature of the Gas Disks} 

The final rotation curves and mid-plane volume densities of the counterrotating
gas disks at the end of simulations G2 and G4 are shown in Fig.~\ref{gasdens}.
The rotation velocity and velocity dispersion values (azimuthally averaged)
are shown in the plots at the top, with the SPH-computed total mid-plane gas
density (azimuthally averaged) at each radial location is shown with filled
circles in the bottom plots.  This mid-plane density is compared to
$\rho_{00}$, the unperturbed central density for the plane-parallel obtained
by adding up the contributions due to the gas pressure and the (external)
gravitational pressure as per equation (9) in \cite{ee78}: {\small\bea
\rho_{00} = ( P_{\rm ext} + \frac{1}{2}\pi G\sigma^2) / c^2. \nonumber\eea}
This ``expected'' mid-plane density is plotted as the open circles, and the
second term on the right of the above equation, due to the gas surface
density, is plotted as the open squares.  The gravitational pressure in the
first term is obtained from the combined halo, primary disk and gas vertical
forces shown in Fig.~\ref{fgrav}.

For the cold primary (G2) gas disk, the rotation velocity is steep initially
and levels off at $R\sim3$ kpc, whereas the hot primary (G4) gas disk's
rotation curve has a more gentle slope towards its flat portion.  This is due
to the higher halo core radius for the NGC~4138 model, which reduces the
halo's influence in the inner few kpc.  The heights of the flat portions of
the rotation curves are comparable for the gas disks in G2 and G4, but the G2
gas disk clearly shows a higher average velocity dispersion.  The
counterrotating disk formed in the cold primary is evidently somewhat hotter
than the one formed in the hot primary.

The value of $\rho_{00}$, the expected mid-plane density obtained by adding up
the two contributions from the gravitational and gas pressures, compares well
with the actual computed mid-plane gas density, although there are significant
differences.  For the cold primary, the expected value is consistently less
than the actual density except for the outer disk.  In the case of the hot
primary, the reverse is true.  The differences seen in the computed and
expected densities are probably due to errors in the surface density and
gravitational pressure values obtained from the particle data.  Sources of
uncertainties include the cutoff height assumed for the gas disk, the
correction applied for the disk center-of-mass and inclination (especially in
G2), and lower particle numbers at higher radii.  The curves for the gas
pressure (surface density) term indicate that the gas pressure is the dominant
contributor for the gas density for the hot primary, whereas for the cold
primary the gravitational pressure is the dominant influence, especially in
the outer disk where the gas surface density falls off considerably.

\ss{Heating of Primary Disk}

The initial and final rotation curves for the primary disks after gas infall
are shown in Fig.~\ref{rcurves} for both the cold (G1-G2) and hot (G3-G4)
primary simulations.  The final rotation velocities and velocity dispersions
are consistent with those observed for S0/Sa galaxies (e.g., \cite{ss96};
\cite{rfi92}) since $v_{\rm rot}/\sigma\lesssim2$ rather than $\gg1$ as is
true for Sb and later types.  Furthermore, it is clear from Fig.~\ref{rcurves}
that {\em both} the cold and hot primaries end up looking like S0s after the
formation of the counterrotating disk.  The velocity dispersion ranges from
$\sim100$ to $\sim150$ km/s for the final disks in either case, with the
rotation velocities dropping by $\sim50-100$ km/s for the cold primary, and
$\sim25-30$ km/s for the hot primary.  In other words, the cold primary fares
much worse than the primary that was hot (S0/Sa) to begin with.  In the same
vein, the outer portions of the primaries, which are colder initially (since
the velocity dispersion is initially proportional to the surface density), are
heated up more than the inner regions.

As mentioned above, the heating seen in simulations G1-G4 is more than was
seen for the corresponding sticky particle simulations, because the infalling
gas columns are shorter in the SPH simulations.  However, the fact that colder
disks are more susceptible to heating is independent of the actual amount of
heating experienced.  Although this vulnerability of cold disks is hardly a
surprise, it does strengthen the argument for predicting that most instances
of massive counterrotating disks are likely to be found in S0/Sa galaxies.
This is consistent with the observations to date.

\ss{Tilting of Disks}
It is apparent from most of the simulations presented above that the resulting
counterrotating gas disk is significantly tilted with respect to the initial
symmetry plane of the primary disk (the $z=0$ plane).  This is not a
phenomenon new to our SPH simulations.  It was observed in the sticky particle
simulations as well and we have commented on it previously (TRJB).  It is
worth reiterating here that the tilting experienced by the gas disk is in
phase with the tilting of the primary stellar disk, occurs over relatively
short time-scales (order of dynamical time of primary disk) compared with the
overall time-scale of each simulation, and is in response to the combined
torques exerted by the halo and the gas.

Even though the net gravitational torques and the angular momentum vectors are
initially all aligned with one another, the forces experienced by individual
particles subsequently are by no means all in the initial plane of the primary
disk.  This is especially true for the gas particles, which undergo gas
dynamical interactions.  For instance in G2, which shows the maximum final
disk inclination, the side views for $t=3.0$ and $t=4.0$ Gyr
(Fig.~\ref{g2gas}) show an excess of gas particles below the disk plane (i.e.,
in the lower half of the panels, which are centered on $z=0$) on the left side
of each panel.  This excess mass concentration below the plane on one side is
consistent with the subsequent tilt of the gas disk (which tilts in phase with
the primary disk) seen in later panels.

Comparisons of the total torques exerted by the halo and the infalling gas on
the primary disk with the rate of change of the angular momentum vector (TRJB)
of the disk show that the two quantities are consistent with each other.  The
change in the value of the $z$-component of the {\em total} angular momentum
vector, which is by far the dominating component, is less than 1\% over the
course of the simulation in most cases, and up to $\sim3$\% in the extreme
cases.

\ss{Star Formation}
Our SPH code does not incorporate star formation as yet.  Previous studies
suggest that in interactions of galaxies where substantial amounts of gas are
involved, the star formation rates are very modest for more than 90\% of the
interaction history, with almost all of the star formation occurring in a rapid
starburst after the gas has dissipated most of its kinetic energy and formed
dense regions in the centers of the galaxies (\cite{mrb92}; \cite{mbr93};
\cite{mh94a}, 1994b).  The pre-starburst phase of the star formation in these
simulations had no discernible effect on the final structure of the gas, since
the depletion rates were very low.  In our case, the longer time scales
involved in our retrograde simulations may deplete significantly more gas over
the course of the simulation, but other than reducing the intensity of the
central starburst, this is unlikely to have a profound effect on the structure
of the gas disks formed as long as there is little or no prograde gas in the
primary disk.  If there is a considerable amount of preexisting prograde gas,
then it is possible that star formation resulting from the head-on collision
of the counterstreaming gas particles will prevent a significant amount of gas
from falling to the center due to rapid conversion to stars.  This may
alleviate the problem of excessive dissipation that causes the gas to collapse
to the center of the primary in our simulations that include prograde gas (G5
and G6).  The presence of star formation may also be instrumental in producing
counterrotating radial profiles that are closer to the primary radial profiles
(i.e., closer to exponential profiles).  We plan to include star formation in
our code in the future, but for now we have excluded it to save precious CPU
time.

\section{Conclusions}

There are some notable differences between the characteristics of the
counterrotating disks resulting from SPH and those obtained with sticky
particle simulations by TR and TRJB.  The SPH disks are very thin compared to
their sticky particle counterparts and compared to the primary disks.  They
also show evidence of spiral structure, and their size and radial mass
distribution are quite sensitive to the input parameters, particularly those
that affect the initial angular momentum of the gas.

Other differences in the SPH results include the lack of clumping of the
infalling gas, a problem that was quite severe with our sticky particle
simulations, and the shorter time-scales for disk formation.

Although it is easy to produce thin counterrotating disks with gas infall, it
is not so easy to obtain exponential radial profiles.  The initial angular
momentum of the gas has to be low enough, and some combination of other
processes such as prograde gas, star formation and energy feedback from
massive stars may be necessary to produce counterrotating disks that are very
similar to the primary disks.  Currently there is no evidence to indicate that
counterrotating disks have predominantly exponential profiles, so this is not
necessarily a problem.

In general, the process that dumps a massive counterrotating disk in a cold
primary spiral, especially if it is a minor gas-rich merger but even if it is
gas infall that occurs over a few dynamical times ($\lesssim10 t_{\rm dyn}$),
is likely to heat up the primary substantially and change its type to an S0/Sa
galaxy.  On the other hand, if the primary is already an S0/Sa galaxy to begin
with, then it can acquire a counterrotating disk without changing its type
significantly.  The fact that most of the currently known instances of massive
spiral counterrotating disks are in S0/Sa galaxies is therefore a selection
effect rather than an accident.

The presence of primordial prograde gas in the primary has a drastic effect on
the retrograde gas that comes in contact with it.  Neutralization of the
angular momenta is rapid, with both the prograde and retrograde gas particles
ending up in the center of the primary within a few dynamical times.  This may
be an indication of a problem with SPH that causes over-dissipation in
counterstreaming gas flows, at least in the absence of star formation.  The
inclusion of star formation and energy feedback from supernovae will most
likely yield significantly different results in such situations.

A retrograde-rotating, infalling gas shell produces a counterrotating bulge
and flat outer ring, but is unable to produce a counterrotating disk in the
proper sense.  The size of the ring is well correlated with the angular
momentum of the shell.  The formation of the ring is consistent with previous
studies of collapsing isothermal gas clouds.  These studies also suggest that
significantly hotter gas with lower angular momentum is necessary to produce a
counterrotating disk with this model.

We hope to test our results further in the near future with the addition of
thermal effects and star formation to our SPH code.

\acknowledgements 

A.R.T. thanks the following people for their help and advice on various
topics: Lars Hernquist (TREESPH), Neal Katz and David Weinberg (TREESPH, time
integration), Rick Pogge (star formation, enrichment and many other issues),
and Chris Mihos (modeling star formation in galaxy interaction simulations,
SPH).  This work was partly supported by a NYI award to B.S.R. (NSF grant
AST-9357396), NASA Grant NAG 5-2864, and Ohio Supercomputer Center grant
PAS825.  A.R.T. is also grateful to Alex Szalay and the Sloan Digital Sky
Survey for support for part of this research.  The simulations were run on the
Cray Y-MP at the Ohio Supercomputer Center, and we thank Barbara Woodall,
David Ennis and Tim Rozmajzl at the OSC for their assistance.  This research
has made extensive use of the NASA ADS Abstract Service and the LANL
Astrophysics E-prints websites.

\clearpage
\begin{figure*}
\caption{Run G1 - counterrotating disk formation by retrograde infall of gas
onto a cold primary disk.  The top (upper 8 panels) and side (lower 8 panels)
views are shown.  The primary disk is shown only in the first panel of each
view for clarity.  The first two panels for each view are 160 kpc wide, with
subsequent panels having the zoom factors indicated.  Time is in Gyr.
\label{g1gas}}
\end{figure*}

\begin{figure*}
\caption{Run G2 - infall of gas with lower angular momentum than in G1.  Panel
widths and time units are the same as in Fig.~\ref{g1gas}.  The radial
distribution of the gas is much closer to an exponential disk rather than a
ring as in G1.
\label{g2gas}}
\end{figure*}

\begin{figure*}
\caption{A comparison of the primary disk for infall simulation G2 at $t =
1.5$ Gyr (top row) and $t = 6.0$ Gyr (middle row) and the counterrotating gas
disk at $t = 6.0$ Gyr (bottom row), with the top views on the left and the
side views on the right.  The width of each panel is 40 kpc.
\label{g2disks}}
\end{figure*}
\begin{figure*}
\caption{Run G2 - velocity fields of the primary disk (left) and
counterrotating gas (right).  Panel width is 40 kpc, and a fraction of the
particles is sampled for clarity, with sampling proportional to radius.
\label{g2vf}}
\end{figure*}

\begin{figure*}
\caption{A comparison of the radial density profiles of the counterrotating
disks for runs G1 (dotted line) and G2 (dashed line), measured at the final
timesteps.  The surface particle number-density $N$ is calculated by counting
and azimuthally averaging the total number of particles in columns of unit
area and half-height (1 kpc$^2$ and 1 kpc respectively), and plotted as a
function of radius $R$.  A correction is applied for the disk inclination if
necessary.  The radial profile of the primary disk at the end of G2 (the G1
primary profile is very similar) is also plotted for reference (solid line).
\label{g12rad}}
\end{figure*}

\begin{figure*}
\caption{A comparison of the primary disk thickness for runs G1 and G2.  The
mean half-thickness $z$ is plotted as a function of radius $R$.  The solid
curve shows the initial primary disk at $t=1.5$ Gyr, and the isolated disk
curve shows the thickness after evolving the primary disk without any gas
accretion for the same time period.
\label{g12zth}}
\end{figure*}
\begin{figure*}
\caption{Run G3 - counterrotating disk formation by retrograde infall of gas
for NGC~4138 primary disk.  The top (upper 8 panels) and side (lower 8 panels)
views are shown.  The primary disk is shown only in the first panel of each
view for clarity.  The first two panels for each view are 80 kpc wide, with
subsequent panels having the zoom factors indicated.  Time is in Gyr.
\label{g3gas}}
\end{figure*}

\begin{figure*}
\caption{Run G3 - velocity fields of the primary disk (left) and
counterrotating gas (right).  Panel width is 17 kpc, and a fraction of the
particles is sampled for clarity, with sampling proportional to radius.
\label{g3vf}}
\end{figure*}

\begin{figure*}
\caption{Run G4 - infall of gas with lower angular momentum and a shorter
column of gas than in G3.  Panel widths and time units are the same as in
Fig.~\ref{g3gas}.  The radial distribution of the gas is much closer to an
exponential disk rather than a ring as in G3.
\label{g4gas}}
\end{figure*}

\begin{figure*}
\caption{A comparison of the radial density profiles of the counterrotating
disks for runs G3 (dotted line) and G4 (dashed line), measured at the final
timesteps.  The surface particle number-density $N$ is calculated as in
Fig.~\ref{g12rad} and plotted as a function of radius $R$.  The radial profile
of the primary disk at the end of G3 (the G4 primary profile is very similar)
is also plotted for reference (solid line).
\label{g34rad}}
\end{figure*}

\begin{figure*}
\caption{Run D1 - gas-rich dwarf merger M2 from TRJB repeated with SPH.  Panel
widths and time units are the same as in Fig.~\ref{g3gas}.
\label{d1gas}}
\end{figure*}

\begin{figure*}
\caption{Run D1 - velocity fields of the primary disk (left) and
counterrotating gas (right).  Panel width is 20 kpc, and a fraction of the
particles is sampled for clarity, with sampling proportional to radius.
\label{d1vf}}
\end{figure*}

\begin{figure*}
\caption{Run D2 - partial SPH run of gas-rich dwarf merger M2 from TRJB.
Panels are 10 kpc wide.
\label{d2gas}}
\end{figure*}

\begin{figure*}
\caption{Run G5 - A prograde ring of gas is included in the primary disk.
Both the prograde and retrograde gas particles are shown.  The primary stellar
disk is not shown.  Panel widths and time units are the same as in
Fig.~\ref{g4gas}.
\label{g5gas}}
\end{figure*}

\begin{figure*}
\caption{Run G5 - Velocity fields of the primary prograde gas (left) and
incoming, initially retrograde gas (right) shown for the first four timesteps
of Fig.~\ref{g5gas}.  Each panel is 4 kpc wide.  Only a fraction of the
particles is shown for clarity.
\label{g5vf}}
\end{figure*}

\begin{figure*}
\caption{Run G6 - A prograde disk of gas is included in the primary disk with
all other input parameters the same as in G4.  Both the prograde and
retrograde gas particles are shown.  The primary stellar disk is not shown.
Panel widths and time units are the same as in Fig.~\ref{g4gas}.
\label{g6gas}}
\end{figure*}

\begin{figure*}
\caption{Run G6 - velocity fields of the corotating (left) and counterrotating
(right) gas at $t=1.1$ (top), $t=1.5$ (middle) and $t=1.7$ (bottom) Gyr.  The
panels on the left are 1.5 kpc wide, and the panels on the right are 4 kpc
wide.  Only a fraction of the gas particles are shown for clarity.
\label{g6vf}}
\end{figure*}

\begin{figure*}
\caption{Run G7 - Retrograde infalling (initially spherical) gas shell.  The
top views (looking along the $z$-axis) are shown on the left and the side
views (looking along the $y$-axis) are shown on the right.  The shell is
initially 10 kpc thick and its initial radius is 30 kpc.  The primary stellar
disk is not shown.  Panel widths for the first two panels in each view are 80
kpc, with subsequent panels having the zoom factors indicated.
\label{g7gas}}
\end{figure*}

\clearpage
\begin{figure*}
\caption{Run G8 - Retrograde infalling shell with half the initial angular
momentum compared to the shell in G7.  The top views are shown on the left and
side views on the right.  The primary stellar disk is not shown.  Panel widths
for the first two panels in each view are 80 kpc, with subsequent panels
having the zoom factors indicated.
\label{g8gas}}
\end{figure*}

\begin{figure*}
\caption{Top and side views of the counterrotating disk formed in G4 showing 
evidence of lopsidedness (top) and warps (bottom).
\label{g4disks}}
\end{figure*}

\begin{figure*}
\caption{The azimuthally averaged gravitational force due to the halo
(circles), disk (squares) and gas (triangles) particles, at points located in
the plane of the disk at the plotted radii, with the component in the plane
shown as the solid line, and the component perpendicular to the plane shown as
the dotted line.  The plots are shown for the initial (top) and final (bottom)
configurations, with results for the cold primary simulation G2 being shown on
the left and the hot primary simulation G4 shown on the right.  The force
values are in simulation units, time in Gyr.}
\label{fgrav}
\end{figure*}

\begin{figure*}
\caption{The final rotation velocity (open circles) and velocity
dispersion (open squares) plots (top), and the mid-plane gas density (bottom)
for the gas disks formed at the end of simulations G2 (left) and G4 (right).
In the gas density plots, the actual average mid-plane gas density computed
using SPH is plotted as the filled circles, whereas $\rho_{00}$, the expected
mid-plane density (see text), is shown as the open circles.  The contribution
to the gas density due to the gas pressure (see text) is plotted as the open
squares.  The density values are in simulation units.}
\label{gasdens}
\end{figure*}

\begin{figure*}
\caption{A comparison of the rotation velocity $v_{\rm rot}$ and
velocity dispersion $\sigma$ as a function of cylindrical radius $R$ of the
initial primary disks (top) with the final primary disks (middle and bottom)
for gas infall simulations with a cold primary (left) and hot primary (right).
The open circles show the mean rotation velocity and the open squares the
velocity dispersion.  The velocity unit is $\sim1000$ km/s.  Both velocities
for the final disks have been corrected for disk inclination, but values for
$R\lesssim0.1R_d$, where $R_d$ is the maximum value of $R$ plotted, are not
reliable.  The filled triangles show the ratio $v_{\rm rot}/\sigma$ (limits
shown on the right axis) as a function of radius.}
\label{rcurves}
\end{figure*}

\end{document}